\documentclass[12pt]{article}
\usepackage{graphicx}
\usepackage{hyperref}
\usepackage{url}

\newcommand\pubnumber{DPF2015-65}
\newcommand\pubdate{\today}

\def\nebraska{Department of Physics and Astronomy\\
University of Nebraska-Lincoln, Lincoln, NE 68588-0299}

\def\Title#1{\begin{center} {\Large #1 } \end{center}}
\def\Author#1{\begin{center}{ \sc #1} \end{center}}
\def\Address#1{\begin{center}{ \it #1} \end{center}}

\newcommand\pubblock{\rightline{\begin{tabular}{l} \pubnumber\\
         \pubdate  \end{tabular}}}
\newenvironment{Abstract}{\begin{quotation}  }{\end{quotation}}
\newenvironment{Presented}{\begin{quotation} \begin{center} 
PRESENTED AT\end{center}\bigskip 
      \begin{center}\begin{large}}{\end{large}\end{center} \end{quotation}}
\def\Acknowledgments{\bigskip  \bigskip \begin{center} \begin{large}
             \bf ACKNOWLEDGMENTS \end{large}\end{center}}

\def\beq{\begin{equation}}
\def\eeq#1{\label{#1}\end{equation}}
\def\eeqn{\end{equation}}

\def\beqa{\begin{eqnarray}}
\def\eeqa#1{\label{#1}\end{eqnarray}}
\def\eeqan{\end{eqnarray}}

\let\bar=\overbar

\def\Dslash{\not{\hbox{\kern-4pt $D$}}}
\def\dslash{\not{\hbox{\kern-2pt $\del$}}}

\def\msb{{\bar{\ssstyle M \kern -1pt S}}}

\begin{document}
\begin{titlepage}
\pubblock

\vfill
\Title{CMS Software and Computing: Ready for Run 2}
\vfill
\Author{Kenneth Bloom}
\Address{\nebraska}
\vfill
\begin{Abstract}
  In Run 1 of the Large Hadron Collider, software and computing was a
  strategic strength of the Compact Muon Solenoid experiment. The timely
  processing of data and simulation samples and the excellent performance
  of the reconstruction algorithms played an important role in the
  preparation of the full suite of searches used for the observation of the
  Higgs boson in 2012. In Run 2, the LHC will run at higher intensities and
  CMS will record data at a higher trigger rate. These new running
  conditions will provide new challenges for the software and computing
  systems. Over the two years of Long Shutdown 1, CMS has built upon the
  successes of Run 1 to improve the software and computing to meet these
  challenges. In this presentation we will describe the new features in
  software and computing that will once again put CMS in a position of
  physics leadership.
\end{Abstract}
\vfill
\begin{Presented}
DPF 2015\\
The Meeting of the American Physical Society\\
Division of Particles and Fields\\
Ann Arbor, Michigan, August 4--8, 2015\\
\end{Presented}
\vfill
\end{titlepage}
\def\thefootnote{\fnsymbol{footnote}}
\setcounter{footnote}{0}

Joe Incandela, the spokesperson of the Compact Muon Solenoid (CMS)
experiment, had a huge smile on his face at the press conference after his
CERN seminar on July 4, 2012.  Why?  It was because CMS software and
computing had enabled the discovery of the Higgs boson.  He had just shown
evidence for Higgs decays to five different final states.  The analyses had
used every drop of data available from the Large Hadron Collider (LHC), and
had all the necessary Monte Carlo samples in place.  This could not be said
of the competitor experiment, which had access to greater computing
resources~\cite{bib:WLCGresources}.  But the competition has hardly stood
still in the time since then, and neither could CMS.  This presentation
describes the changes that CMS made in its software and computing tools in
during Long Shutdown~1, and how they have made CMS ready for Run~2.

Run~2 has brought us to a new energy domain, with $\sqrt{s} = 13$~TeV, and
even during 2015 the LHC is delivering enough integrated luminosity to do
real physics with, and perhaps to make a discovery if Nature is kind.
Thus, CMS had to be ready from the start to do all the computing necessary
for the physics.  But the computing requirements are
substantially larger than those of Run~1.  CMS will record data at a rate
of 1~kHz, a factor of 2.5 greater than in Run~1, with larger pileup rates.
To process the data without any improvements to the software would require
an increase in CPU resources by a factor of six.  Thus, new approaches were
needed to address the computational challenges.  The result is a system of
increased agility and flexibility that will enable physics discovery, all
built off the extremely successful systems of Run~1.

\begin{figure}[htb]
\centering
\includegraphics[height=4in]{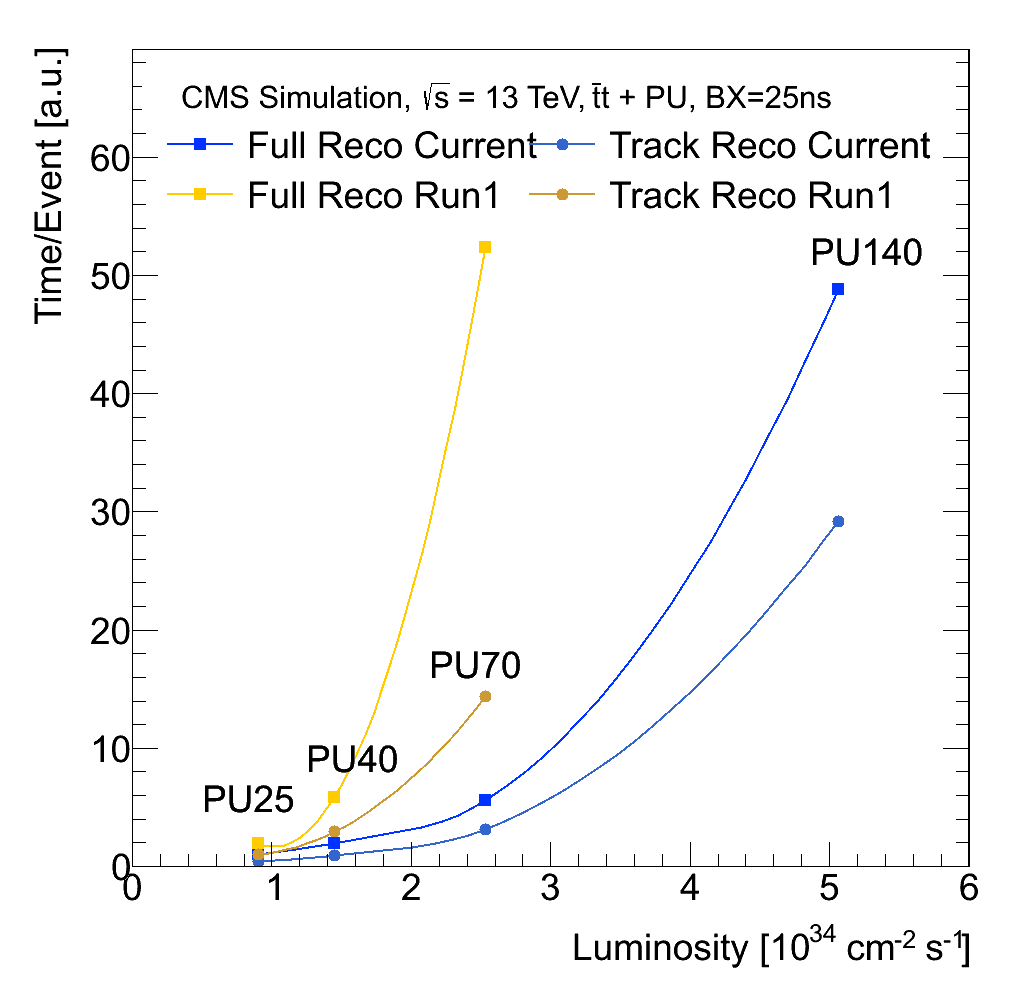}
\caption{Reconstruction time for a $\bar{t}t$ event as a function of
  instantaneous luminosity for two different versions of CMS reconstruction
  software.  The time for track reconstruction only is broken out separately.}
\label{fig:recotime}
\end{figure}

Clearly an important first step was to improve the performance of the
reconstruction algorithms.  As shown in Figure~\ref{fig:recotime}, CMS has reduced the
event reconstruction time while maintaining physics performance, even in
the more difficult event environment of Run~2.  The most important
improvements were in the track reconstruction, which takes the bulk of the
processing time.  Some improvements were strictly techncial, while others
were algorithmic, such as changes to the tracking algorithms that reduced
the number of fake tracks and sped execution.  In addition, the time to
simulate events was improved by reducing the time spent tracking low-energy
particles in {\tt GEANT4}.

Meanwhile, the best thing that the sites in the CMS distributed computing
infrastructure -- seven Tier-1 sites and approximately 50 Tier-2 sites --
could do was to keep running, to maintain operational readiness.
Throughout 2014, there were an average of 18,500 and 54,300 jobs running at
the Tier-1 and Tier-2 sites, respectively.  These jobs allowed CMS
physicists to finish Run~1 data analyses, study detector configurations for
HL-LHC upgrades, and start generating simulation samples for Run~2.

\begin{figure}[htb]
\centering
\includegraphics[height=3in]{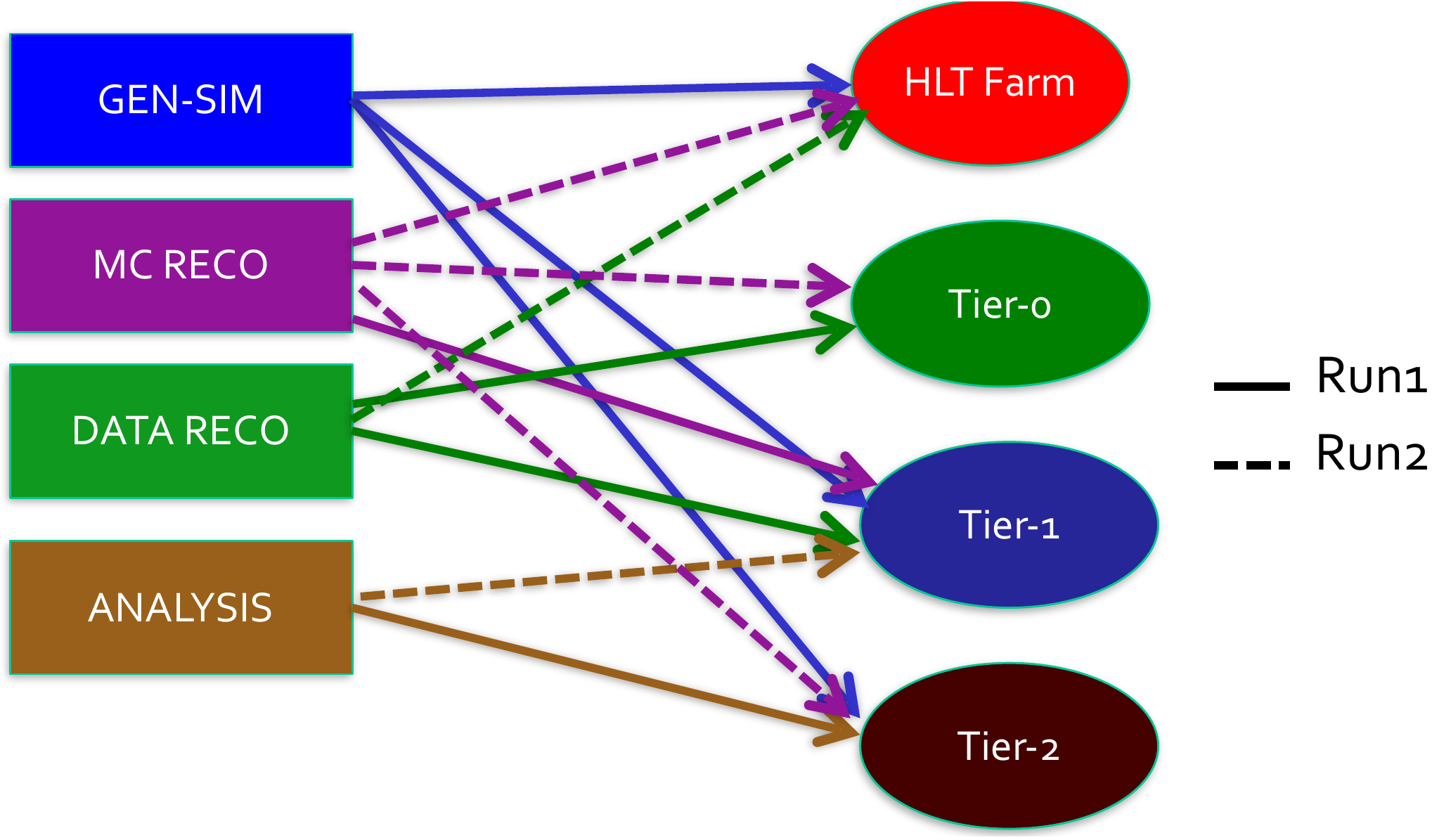}
\caption{Various workflows (in boxes) were run at certain facilities
  (ovals) during Run~1, as indicated by the solid arrows.  The dashed
  arrows indicate the additional workflow-to-facilities mappings that will
  be possible in Run~2.}
\label{fig:workflows}
\end{figure}

One of the goals of the CMS computing organization for Run~2 was to use
these excellent facilities in more flexible and heterogeneous ways.  This
is illustrated in Figure~\ref{fig:workflows}, which shows which workflows
were performed at which types of facilities in Run~1, and how they will be
able to be performed at a greater variety of facilities in Run~2.  The
high-level trigger (HLT) farm, which is approximately the size of the
Tier-0 facility, can now be used for organized processing during technical
stops.  Tier-2 centers have been commissioned to do reconstruction tasks
that were previously limited to Tier-1 sites.  Analysis jobs will now be
permitted to run at Tier-1 sites, with mechanisms in place to ensure that
desired datasets are resident on disk and not only on tape.  The more
places that work can run, the faster the work will go.

Many new services were deployed to support more agile operations.  The
``Any Data, Anytime, Anywhere'' (AAA) global data federation allows CMS
applications to read data efficiently over wide-area
networks~\cite{bib:AAA}.  This relaxes constraints on the locations of
datasets and workflows, and provides a great convenience to physicists who
want to analyze data that isn't resident on their local computer.
Disk-tape separation at Tier-1 sites gives operators greater control over
what datasets are available on disk and, through AAA, allows Tier-1 data to
be used in workflows anywhere.  A new dynamic data management system
performs automatic transfers of datasets upon their creation, and then
deletes them when they are not needed, using ``popularity'' data on file
access rates.  There is now a global job pool for resource provisioning
through the glideinWMS system~\cite{bib:gWMS}.  This single task queue for
all jobs within the distributed computing infrastructure allows for central
control of job priorities, and is simpler to manage than the multiple
provisioning systems used in Run~1.  It has been demonstrated to scale
sufficiently well to operate all Tier-1, Tier-2 and opportunistic resources
in a single pool.  The glideinWMS system also gives the possibility of
provisioning cloud infrastructures.  This allows the use of the HLT and
possible opportunistic and commercial cloud resources, and the ability to
burst into extra resources when necessary.  Finally, data access across the
ocean is facilitated by the establishment of a 100~Gbps transatlantic
network link that is provided by ESnet.

A number of advances within the CMS software have also provided new
opportunities and efficiencies.  The event-processing framework and the
reconstruction code are now multi-threaded.  The code can use several CPU
cores concurrently to reconstruct multiple events simultaneously.  This has
several beneficial effects.  There is less demand on the computing
infrastructure in general, as processing can be done with fewer open files
and fewer jobs.  The time to process a luminosity block of data (the basic
quantum of detector data) is reduced, which is needed for higher trigger
rates.  There is also a huge reduction in the memory required per CPU core
with little efficiency loss in the throughput of event processing.  The
multi-threading enables the use of multi-core pilot jobs that have internal
dynamic partitioning of resources for greater efficiency.

CMS has also provided better tools for physics users, making data analysis
easier, more flexible and less resource-intensive.  A new job submission
tool, CRAB3, has many improved features over its predecessor.  There are
automatic job retries, better job tracking, and more reliable delivery of
user output.  CRAB3 has a thinner client layer, and more logic on the
server side that allows for easier upgrades.  The tool fully exploits the
HTCondor~\cite{bib:htcondor} and glideinWMS systems, including overflowing
jobs from busy to less-busy sites.  The new {\tt miniAOD} analysis data
format is a mere 30~KB/event, one tenth the size of the current {\tt AOD}
format.  It is designed to serve about 80\% of analyses.  A format this
small makes it easier to keep more of the data disk-resident at desired
locations.  Once again, the AAA federation plays a role as user job
location is no longer tied to data location.  This is a major enabler for
university-based data analyses, as smaller university groups are not always
able to maintain large storage systems.

All of the improvements discussed are targeted for LHC Run~2, but CMS must
always be preparing for the future.  All software and computing systems
have natural life cycles that require regular re-engineering, to continue
coping with increasing LHC requirements, to adapt to changing technologies,
and to improve usability and maintainability.  Unlike detector upgrades,
the deployment of new systems is not necessarily coupled to the LHC
operation schedule, which means that they can have a different development
and deployment cycles than instrumentation.  A current hot project is the
use of large-scale commercial cloud resources.  The goal is to be able to
rapidly expand resources for burst needs.  CMS is working with Amazon Web
Services on a demonstration project to increment the experiment's computing
resources by about 50\% for an expected fall reprocessing campaign.  If
such demonstrations are successful (a big ``if''), it might become possible
to only own the processing resources required for average demand rather
than peak demand.  Other ongoing projects include developing tools,
environments and algorithms needed for new architectures, such as GPU's and
low-energy clusters, and exploring the evolution of computing models and
the scaling of tools for LHC Run~3.

How is the new software and computing system performing in the beginning of
Run~2?  There is not much data yet, so the system has not really been
tested at scale, but everything has worked fine so far.  The Tier-0
facility does prompt reconstruction of events, and physicists soon have the
data in hand.  The first 13~TeV physics paper, on $dN/d\eta$, has been
submitted for publication~\cite{bib:dndeta}.  Other preliminary results
available at this time include a first measurement of the top-antitop
production cross section~\cite{bib:ttbar} and a measurement of the dijet
mass spectrum~\cite{bib:dijet}.  Obviously, much more excitement is
expected, and CMS software and computing tools are needed to make all of
these measurements possible.

While CMS software and computing were very successful in Run~1, the
experiment could not -- and did not -- sit still during the long shutdown.
Significant evolutionary changes to the Run~1 systems have taken advantage
of technical developments that have led to more
flexible and efficient resource usage and better tools for physics users.
CMS is already showing significant success in the processing and analysis
of Run~2 data.  If Nature cooperates, CMS software and computing will have
everyone smiling again.

\Acknowledgments
I thank my CMS colleagues for all of their work to build this successful
software and computing system, and especially my fellow leaders of the
U.S. CMS Software and Computing Operations Program who have given me
feedback over time on predecessors to this presentation.  My work on this 
project is supported by grants that include NSF-1120138 and NSF-1306040.

\end{document}